\documentclass[twoside,showpacs,superscriptaddress,twocolumn,floatfix,a4paper,aps]{revtex4}
\usepackage{color}
\usepackage{amsmath} 
\usepackage{graphicx} 
\usepackage[utf8x]{inputenc}
\usepackage[T1]{fontenc}

\newcommand{\etal}{\textit{et al.}}
\newcommand\extra[1]{}
\newcommand\Ns[2]{\left(
    \begin{array}{c}
      #1 \\
      #2
    \end{array}
  \right)}
\newcommand\mmax{m_f } 
\newcommand\mmin{m_i } 

\renewcommand\emph[1]{{\it #1}}
\renewcommand\em{\it}

\begin{document}

\title{Using quantum routers to implement quantum message authentication\\ and Bell-state manipulation}

\author{Karol Bartkiewicz} \email{bartkiewicz@jointlab.upol.cz}
\affiliation{Faculty of Physics, Adam Mickiewicz University,
PL-61-614 Pozna\'n, Poland}
\affiliation{RCPTM, Joint Laboratory of Optics of Palacký University and Institute of Physics of Academy of Sciences of the Czech Republic, 17. listopadu 12, 771 46 Olomouc, Czech Republic}
\author{Antonín Černoch} \email{cernoch@jointlab.upol.cz}
\affiliation{Institute of Physics of Academy of Sciences of the Czech Republic, Joint Laboratory of Optics of PU and IP AS CR, 
   17. listopadu 50A, 772 07 Olomouc, Czech Republic}
\author{Karel Lemr}
\email{k.lemr@upol.cz}
\affiliation{RCPTM, Joint Laboratory of Optics of Palacký University and Institute of Physics of Academy of Sciences of the Czech Republic, 17. listopadu 12, 771 46 Olomouc, Czech Republic}

\date{\today}

\begin{abstract}
In this paper we investigate the capability of quantum routing (quantum state fusion) to implement two useful quantum communications protocols. The analyzed protocols include quantum authentication of quantum messages and non-destructive linear-optical Bell state manipulation. We also present the concept of quantum decoupler -- a device implementing an inverse operation to quantum routing. We demonstrate that both quantum router and decoupler can work as specialized disentangling gates. 
\end{abstract}


\maketitle
\section{Introduction}
Development of future quantum communications networks {relies} on successful implementation of several key protocols \cite{Nielsen_QCQI,Zeilinger_QIP,bib:bruss:quantum_information}. These protocols are  quantum analogues {of} their classical counterparts such as amplification \cite{gisin10ampl,Pitkanen11ampl,Curty11ampl,micuda12ampl,Kocsis13NPhys9,bula13qnd,scott13ampl}, secure key distribution \cite{Gisin02crypto,Bartkiewicz13,Ursin07QKD,Wang12QKD-260km,Makarov10Hacking,Lo12MDI-QKD} or error correction \cite{shor95error,calderbank96error,steane96error,gottesman98error,modlawska:non_maxi_entangl,paetznick13error,weinstein13error}. Another active element needed for {the} construction of complex quantum networks is quantum router -- a quantum-mechanical {counterpart} of the classical router used to steer the information from its source to intended destination. This component has been subject of both theoretical and experimental research. In some cases, classical information is used to control the path of quantum signal \cite{Hall11}, {however other implementations} can be called fully quantum routers since both the control and signal are quantum. In our recent papers, we have presented linear-optical schemes of quantum routers together with several criteria the router has to fulfill in order to be fully functional \cite{lemr12router,lemr13router}. A similar quantum information protocol entitled {\it quantum state fusion} has been experimentally implemented by Vitelli {\it et al.}  \cite{qfusion13}  independently of the above-mentioned studies.  Other physical systems, {e.g. light-atom interaction \cite{Zueco09,Aoki09,Hoi11,liao10router,zhou13router,lu14router},} have also been considered for construction of quantum routers.

However, the quantum router has {so far} only been considered a mere replacement of the classical router for the purposes of quantum networks. In this paper, we go beyond this analogy and {discuss} the idea of using the quantum router to implement {other} useful quantum communications protocols. The quantum router will play a key role in complex quantum communications networks. Therefore, investigating its potential applications beyond the simple routing is of great significance.

{The article is organized as follows.} In Sec. \ref{ref:qr}, we review the conceptual scheme of a quantum router and present its mathematical description. We also present the concept of a {\it quantum decoupler}, which is an inverse component to the quantum router. Specific linear-optical implementations of both the router and the decoupler can be found in Refs.~\cite{lemr12router,lemr13router,qfusion13}. The following Sec.~\ref{ref:sign} describes a proposal for quantum authentication of quantum messages using the quantum router and decoupler. Finally, in Sec. \ref{sec:bell} we present a protocol for {nondestructive} Bell state manipulation using the quantum router. We conclude in Sec.~\ref{sec:concl}.

\section{Quantum routing and decoupling}
\label{ref:qr}
\begin{figure}
\includegraphics[scale=1]{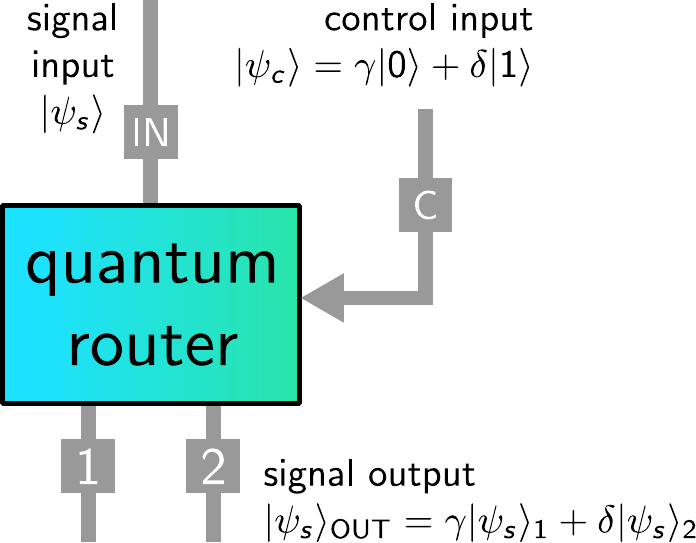}
\caption{\label{fig:concept:router}(Color online) Conceptual scheme of a quantum router as described in the text. Signal qubit $|\psi_s\rangle$ is routed into a coherent superposition of two output modes depending on the state of the control qubit $|\psi_c\rangle$.}
\end{figure}
\subsection{Quantum routing}
Quantum routing (or quantum state fusion \cite{qfusion13}) is an essential protocol allowing for construction of complex quantum networks. In contrast to a classical router, the quantum router {exploits quantum phenomena}. Thus for instance the multi-direction broadcasting (sending multiple copies of the signal to all output ports) known from classical routing is forbidden since it would contradict the no-cloning theorem. On the other hand, various quantum {phenomena} such as {quantum superposition} or entanglement can be used. In our recent paper, we have defined a set of requirements imposed on a fully functional quantum router \cite{lemr13router}. These requirements include capability of the router to {split} quantum signal {depending on the instructions passed by the control qubit while keeping the signal qubit intact}. The basic one-to-two port quantum router is schematically depicted in Fig. \ref{fig:concept:router}. Assuming an initial signal qubit encoded in state $|\psi_s\rangle$ and control qubit {reads}
\begin{equation}
|\psi_c\rangle = \gamma|0\rangle + \delta|1\rangle.
\end{equation}
The quantum router transforms the signal state $|\psi_s\rangle$ to 
\begin{equation}
\label{eq:router}
|\psi_s\rangle_\mathrm{OUT} = \gamma|\psi_s\rangle_1 + \delta|\psi_s\rangle_2,
\end{equation}
where indices 1 and 2 denote output signal modes. Note that the signal state remains intact.

Let us consider single photon qubits in polarization and spatial mode encoding. In this case, the signal qubit is encoded in polarization {degree degree of freedom as} 

\begin{equation}
|\psi_s\rangle = \alpha|H\rangle + \beta|V\rangle,
\end{equation}
where $H$ and $V$ denote horizontal and vertical polarizations{, respectively}. The control qubit is stored in another polarization encoded photon state

\begin{equation}
|\psi_c\rangle = \gamma|H\rangle + \delta|V\rangle.
\end{equation}
At the output, we find the signal photon leaving in superposition of both output ports depending on the control qubit {$|\psi_c\rangle$} while keeping the original signal qubit {$|\psi_s\rangle$} undisturbed. {The state of the output is given as}
$$
|\psi_s\rangle_\mathrm{OUT} = \gamma\left(\alpha|H\rangle + \beta|V\rangle\right)_1+ \delta\left(\alpha|H\rangle + \beta|V\rangle\right)_2,
$$
{where indices 1 and 2 denote spacial modes.}

\begin{figure}
\includegraphics[scale=1]{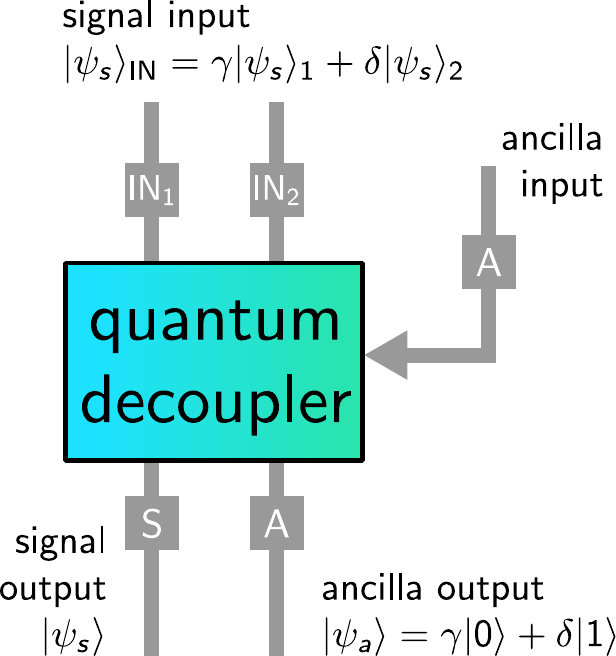}
\caption{\label{fig:concept:decoupler}(Color online) Conceptual scheme of a quantum decoupler -- a device inverse to a quantum router. {There are two qubits stored in the signal photon. The first qubit is encoded in its spatial degree of freedom and the second in polarization}. Quantum decoupler facilitates interaction between this signal input state and an ancillary qubit leading to transcription of the spatial mode encoded qubit to the state of the ancillary qubit while the second signal qubit remains stored in the signal state.}
\end{figure}

\subsection{Quantum decoupling}
One can also investigate inverse procedure to the routing transformation {or quantum synthesis}. In this case, two separate qubits are decoupled from an initial state {expressed by Eq.~}(\ref{eq:router}). {We will refer to this procedure as to quantum decoupling or quantum decomposition.} The transformation implemented by {the} decoupler reads
\begin{equation}
\label{eq:decoupler}
|\psi_s\rangle_{\mathrm{IN}}=\gamma|\psi_s\rangle_1 + \delta|\psi_s\rangle_2 \rightarrow |\psi_s\rangle|\psi_a\rangle,
\end{equation}
where $|\psi_a\rangle$ is the state of ancillary qubit $|\psi_a\rangle = \gamma|0\rangle + \delta|1\rangle$. Conceptual scheme of such decoupler {is shown} in Fig. \ref{fig:concept:decoupler}.

{In the analyzed case of} linear-optical implementation with polarization and spatial mode encoding {we express} the decoupler transformation in terms of horizontal and vertical polarizations and spatial modes
\begin{eqnarray}
|\psi_s\rangle_{\mathrm{IN}}&=&\gamma\left(\alpha|H\rangle + \beta|V\rangle\right)_1 + \delta\left(\alpha|H\rangle + \beta|V\rangle\right)_2\nonumber\\ &\to &
\left(\alpha|H\rangle + \beta|V\rangle\right)_S\otimes\left(\gamma|H\rangle + \delta|V\rangle\right)_A,
\end{eqnarray}
where {$S$} and {$A$} stand for signal and ancillary modes{, respectively}.

\subsection{Polarization -- spatial mode encoding swap}
\begin{figure}
\includegraphics[scale=1]{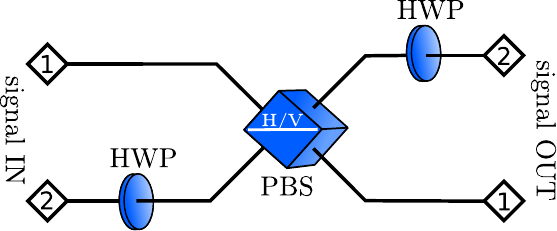}
\caption{\label{fig:swap} (Color online) Scheme for polarization -- spatial encoding swap{: PBS,  fully polarizing beam splitter (transmitting horizontal polarization and reflecting vertial polarization); HWP,}  half-wave plate (rotated by 45 deg. with respect to horizontal polarization).}
\end{figure}
Some applications (for instance the quantum authentication protocol described in Sec. \ref{ref:sign}) require the capability to swap spatially and polarization encoded qubits before the signal is processed by the quantum decoupler. This task can easily be achieved deterministically using the setup depicted in Fig.~\ref{fig:swap}. It consists  of two half-wave plates rotated by 45 deg. with respect to {the} horizontal polarization, and one fully polarizing beam splitter transmitting horizontal polarization and reflecting vertical polarization. In terms of signal state transformation, this component transforms
\begin{equation}
|\psi_s\rangle = \gamma\left(\alpha |H\rangle + \beta |V\rangle\right)_1 + \delta\left(\alpha |H\rangle + \beta |V\rangle\right)_2
\end{equation}
{into}
\begin{equation}
|\psi_s\rangle' = \alpha\left(\gamma |H\rangle + \delta|V\rangle\right)_1 + \beta\left(\gamma |H\rangle + \delta |V\rangle\right)_2.
\end{equation}
This operation when used for swapping the encoding of the signal input qubits before the decoupler is equivalent to swapping the signal and ancilla outputs of the decoupler.

\section{Quantum authentication of a quantum message}
\label{ref:sign}

In both classical and quantum information research, the problem of message authentication is a heavily investigated topic~\cite{rivest90}. The goal of all these protocols is to provide means to verify the authenticity of message origins and thus protect message from both the falsification and repudiation. In 2001, Gottesman and Chuang patented a scheme that used quantum information to securely sign classical messages~\cite{chuang2002patent,chuang2001sign}. In another seminal paper \cite{barnum02existence} published one year later, the authors have presented a proof that it is impossible to quantum sign quantum messages. Although it is indeed impossible to implement (non-arbitrated) quantum signing of quantum messages, it is still possible to accomplish a closely related task – the quantum authentication of quantum messages. In case of signature protocols the message needs to be readable for everyone, in authentication schemes this condition is relaxed and only a trusted party is able to read the message. In contrast to signing, an authenticated message can be verified through a (secret) key shared between the issuer of the message and its intended recipient. As the authors of Ref.~\cite{barnum02existence} show, authenticated quantum messages must be encrypted, while no such requirement is imposed for classical messages. 

The above mentioned research sparked great interest of the scientific community. Quantum signing of classical messages (classical bits) has been studied for instance in Refs.~\cite{xiao06notrusted,wang06single,sousa06contract}. In 2013, Li~\etal explicitly showed \cite{li13quantum} that quantum signature of a quantum message (qubits) is possible in arbitrated schemes. They also showed that this fact  does not contradict the impossibility of non-arbitrated quantum signing mentioned before \cite{barnum02existence}. A number of protocols implementing this task have been derived so far \cite{lu05fourier, li09arbitrBell, gao11arbitrated, zou10analysis, zeng02arbitrated, curty08optim, zeng08arbutratedqs, zou10qs}.

Apart from quantum signing, attention was also focused on closely related protocols, such as quantum authentication of quantum messages~\cite{barnum02existence,curty02qubitauth}, quantum identification  \cite{dusek99indent}  or fingerprinting~\cite{buhrman01finger}. The potential for future realistic deployment of quantum signature and authentication can be demonstrated on the basis of two recent experiments~\cite{clarke12exper,collins13exper}. 

Since future quantum networks should be able to benefit from all aspects of quantum information, we will focus on protocols involving a genuine quantum message. A number of possible general approaches has been proposed in this matter. For instance, a complex multiparticle GHZ entangled state distributed to the communicating parties can be used for {message signing} \cite{zeng08arbutratedqs}. Its generation and distribution might{, consequently,} prove to be experimentally difficult. {But other} protocols using only Bell states \cite{li09arbitrBell} or separable states \cite{zou10qs} have been also derived. In these cases however, the {sender} has to {have} multiple copies of the quantum message. This might be problematic {because in principle}  quantum information {cannot} be perfectly copied. Two alternative approaches have been proposed by Barnum \etal~\cite{barnum02existence}. These authors overcome the need for multiple copies of quantum message by inserting additional qubits into the data stream and subsequently perform authenticity verification based on syndrome measurement of a stabilizer purity test code. They have derived two schemes. The first one uses a distribution of entangled Bell states, whereas the second only transmits separable states. There is, however, an inherent problem with the linear-optical implementation of these two schemes. In order to perform a general stabilizer purity test, all transmitted qubits have to be collectively processed by a complex multi-qubit quantum circuit. Considering the probabilistic nature of linear-optical quantum gates, this procedure will be highly ineffective and would require additional resources in the form of ancillary photons so that individual quantum gates can be joined together \cite{bula13qnd}.

In this section we propose a feasible quantum authentication protocol for quantum messages that does not rely on entanglement distribution, availability of multiple copies of the message or complex multi-qubit measurements and is, therefore, suitable for the platform of linear-optics. At the same time, our scheme is efficient in the sense that only $n$ photons (each encoding 2 qubits) are necessary to send a $n$-qubit long quantum message. The general idea behind our scheme is to use two degrees of freedom of light, i.e., one degree of freedom for the message itself and another degree of freedom  for a one-time quantum authentication key. If one can decouple the message from the one-time key, one can also attack the scheme by coupling a counterfeited message with {a} previously decoupled key. There are two possible solutions: (i) the key reflects the message (in classical computing the value of an ``irreversible'' function of the message is used as a key), (ii) the message is encrypted in a way that allows only the recipient to verify its authenticity.  The first solution, used for instance in \cite{zou10qs}, {requires} some information about the message {that cannot be provided assuming that no perfect copies of the message exist}. Therefore, we resort to the second approach. This approach, as proved in \cite{barnum02existence}, requires sharing a secret key between communicating parties and some sort of encryption of the transmitted message.  {In our solution, the degree of freedom encoding message or key qubit is chosen at random and this random choice has a similar purpose as message encryption in \cite{barnum02existence}.}

Our authentication protocol follows several steps: Assume Alice wishes to transmit a quantum message $|M\rangle$ composed of $n$ independent qubits $|m\rangle$. These qubits are stored in polarization-encoded states of single photons $|m\rangle = \alpha|H\rangle + \beta |V\rangle$. Her goal is to transmit this message to Bob using a quantum authentication protocol so that Bob can be sure of its authenticity.

{\bf Step 1:} Bob and Alice establish a standard quantum cryptography communication channel such as BB84 or R04. Using this channel, Bob sends two encrypted strings {$\mathcal{S}_1$} and $\mathcal{S}_2$ of classical information each $n$-bit long (referred to as first and second ``salt'' string). This approach is inspired by Assis {\em at al.} \cite{assis12sign}. Alice generates a random one-time key $|K\rangle$ composed of $n$ polarization-encoded qubits $|k\rangle$. She uses the first salt string $\mathcal{S}_1$ to establish the state preparation basis. If the corresponding bit equals 0, she randomly prepares either diagonal {$|+\rangle$} or anti-diagonal {$|-\rangle$} {polarization} states \footnote{Throughout this paper, we adopt the following definition of polarization qubit states: $|H\rangle $ -- horizontal polarization state, $|V\rangle $ -- vertical polarization state, $|+\rangle = \frac{1}{\sqrt{2}}\left(|H\rangle+|V\rangle\right)$ -- diagonal polarization state, $|-\rangle = \frac{1}{\sqrt{2}}\left(|H\rangle-|V\rangle\right)$ -- anti-diagonal polarization state, $|R\rangle = \frac{1}{\sqrt{2}}\left(|H\rangle+i|V\rangle\right)$ -- right circular polarization state, $|L\rangle = \frac{1}{\sqrt{2}}\left(|H\rangle-i|V\rangle\right)$ -- left circular polarization state.}. If the salt bit equals 1, she randomly chooses right ($|R\rangle$) or left ($|L\rangle$) circular polarization states.

{\bf Step 2:} Alice takes first qubit of the message ($|m\rangle$) and {the} first qubit of the key ($|k\rangle$). These two qubits are subsequently inserted into a quantum router. The router encodes the control qubit into the spatial mode of a signal photon while preserving the message in its polarization state. For reference see \cite{lemr13router}.  If the first bit of Bob's second salt string $\mathcal{S}_2$ is 0, then $|m\rangle$ serves as signal information and $|k\rangle$ is the control qubit. Then, the router output state becomes
\begin{equation}
\label{eq:sign_0}
\tfrac{1}{\sqrt{2}}\left[\left(\alpha |H\rangle + \beta |V\rangle\right)_1 \pm \mathrm{i}^r \left(\alpha |H\rangle + \beta |V\rangle\right)_2\right],
\end{equation}
where the $\pm$ sign depends on Alice's choice and the value of $r$ corresponds to the value of the relevant bit in the first salt string.
On the other hand, if the first qubit of Bob's second salt is 1, then the roles of $|m\rangle$ and $|k\rangle$ are reversed ($|k\rangle$ becomes signal and $|m\rangle$ becomes control). This swap leads to the output state
\begin{equation}
\label{eq:sign_1}
\tfrac{1}{\sqrt{2}}\left[\alpha\left(|H\rangle \pm \mathrm{i}^r |V\rangle\right)_1 + \beta \left( |H\rangle \pm \mathrm{i}^r  |V\rangle\right)_2\right].
\end{equation}
Alice then sends the output qubit  encoded both in polarization and spatial mode to Bob.

{\bf Step 3:} Bob receives the photon from Alice and, depending on the value of his second salt string $\mathcal{S}_2$ bit, he either performs an encoding swap  or not . 

This is a simple encoding swap operation that transforms (\ref{eq:sign_1}) into (\ref{eq:sign_0}) and vice versa. Subsequently, Bob subjects the  photon to the decoupling procedure described in Section~\ref{ref:qr}. As demonstrated in that section, Bob obtains the message qubit from the ancillary output port of the decoupling device while he simultaneously performs {a} projection measurement either on {$|+\rangle$} and {$|-\rangle$} or $|R\rangle$ and $|L\rangle$ polarizations in key output port depending on the value of the respective bit of the first salt string $\mathcal{S}_1$.

{\bf Step 4:} After Bob's successful detection, he asks Alice to publish her choice of key state $|k\rangle$ using a quantum authenticated classical channel and compares that to his measurement outcome. 

All subsequent message qubits are transmitted in the same way. It is imperative that Alice uses her authentication key $|K\rangle$ only once.

Note that the above mentioned protocol can be implemented even without a quantum router. One can send the key and message qubits separately. Because the second salt string is used, the order in which they are sent is randomized. On the other hand, the quantum router permits the encoding of both qubits into one physical photon, making the transmission more effective. To demonstrate this effect, consider a transmission line with transmissivity $\tau$. A state of two consecutive photons is then transformed according to 
$$
|km\rangle \rightarrow \tau^2 |km\rangle 
$$
where $|k\rangle$ and $|m\rangle$ denote key and message qubits stored in individual photons. Since the protocol succeeds only if both the photons reach Bob's end, we have post-selected only this case. It occurs with success probability of  $\tau^4$. On the other hand, if one uses the router and thus encodes the information into both polarization and spatial degrees of freedom of one single photon, the state can be expressed in the form of (\ref{eq:sign_0}). Without loss of generality, let us consider the key qubit being in the state $|+\rangle$ and being encoded into the spatial mode. The transformation of this state imposed by the lossy channel now reads
$$
\frac{1}{\sqrt{2}}\left(|m0\rangle + |0m\rangle\right) \rightarrow \tau \frac{1}{\sqrt{2}}\left(|m0\rangle + |0m\rangle\right).
$$
It follows from this equation that the success probability (the probability of successful transport) is now $\tau^2$ which is a significant improvement over the non-router based strategy especially for small channel transmissivities.

\subsection*{Message falsification attempt}
Let us begin our analysis by considering that Eve has no {\it a priori} information about the message sent by Alice to Bob. If Eve attempts to falsify it (replace it with her own message), she has to guess which of the two possibilities (\ref{eq:sign_0}) or (\ref{eq:sign_1}) has been selected (because she does not know the salt string). If she guesses {correctly} (in half of the cases), she can falsify the message qubit. {If she guesses incorrectly}, she {will mistakenly use} the message $|m\rangle$ as the key. In {this} case, there is a $\frac{1}{2}$ possibility that Bob {will} detect an error ({assuming} the message is random). Therefore, Eve trying to falsify the information introduces  an inherent error rate $\epsilon = \frac{1}{4}$. For $n$ number of message qubits, the probability of Eve managing to counterfeit $n$ qubits and not be detected is given by
\begin{equation}
P_C = \left(1-\epsilon\right)^n = \left(\frac{3}{4}\right)^n.
\end{equation}
For example{, if  $n$ equals 10 it is} about 6\%. For $n$ equal to 20, it would be as low as 0.3\%. 
A problem can arise when the message is too short, or if Eve only tries to counterfeit a small fraction of the message. To resolve this issue Alice and Bob can introduce decoy states. As a decoy state Alice sends {an} orthogonal state to the current key state instead of {the} real message qubit. After Bob receives this state, she informs him that it was a decoy and Bob performs verification measurement. In this case even if Eve guesses correctly the salt string, there is a $\frac{1}{2}$ probability that Bob will detect an error by measuring polarization of the message qubit replaced by Eve with falsified data. The overall probability of successful counterfeiting $k$ qubits from a $(n+d)$-qubit string  is now given as
\begin{equation}
\label{eq:PC}
P_C(n,d,k,m) = \frac{3^k\Ns{d}{m} \Ns{n}{k-m}}{4^k2^m\Ns{n+d}{k}} 
\end{equation}
where  $d$ is the number of randomly inserted decoy states and $m$ is the number of decoy states attacked. All qubits are successfully counterfeited with probability  $P_C(n,d,n+d,d) = (\tfrac{3}{4})^{n}(\tfrac{3}{8})^d$. This probability drops rapidly with the length of the transmitted string.

\begin{figure}
\includegraphics[width=8cm]{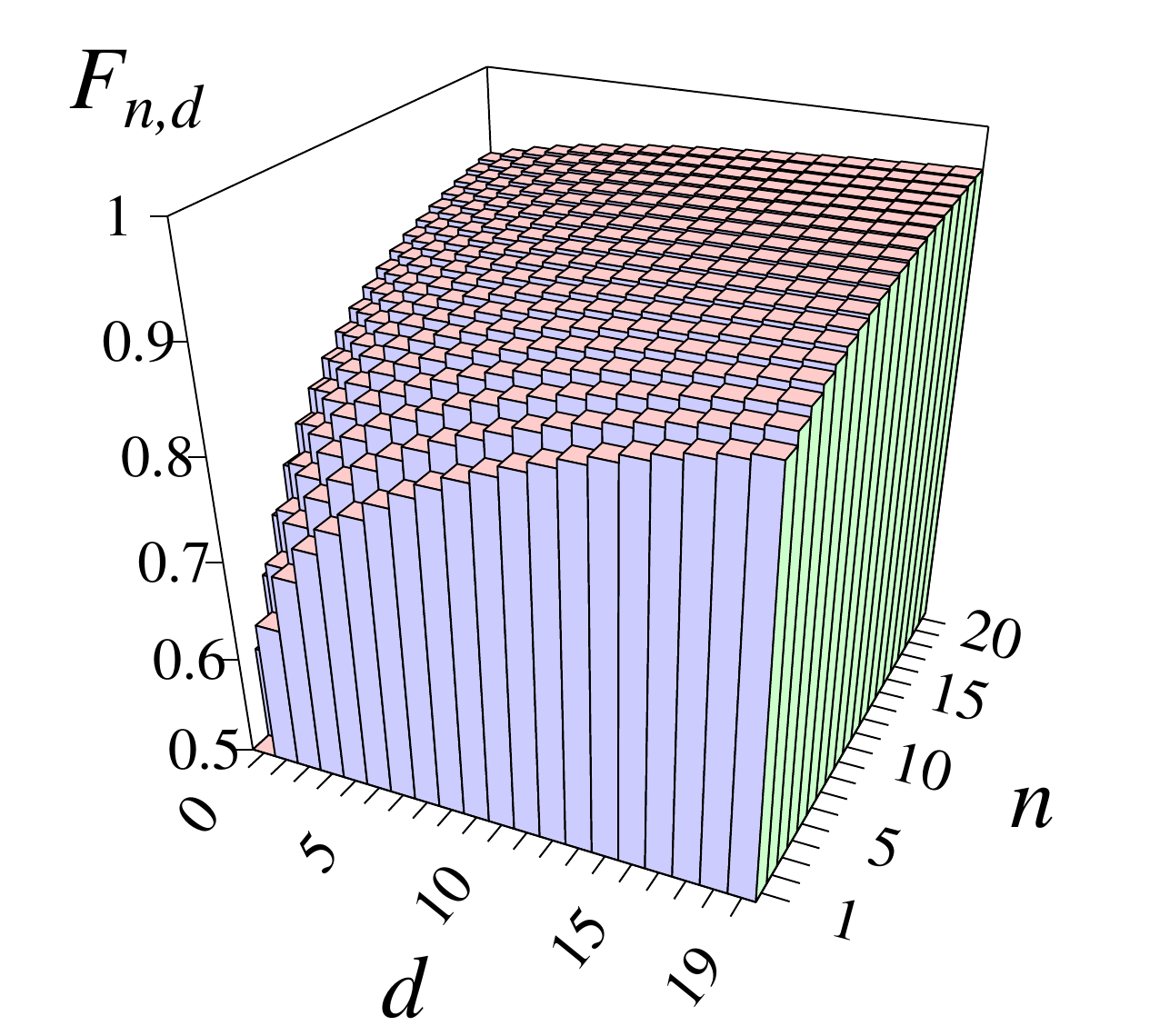}
\caption{\label{fig:Fnd} (color online) The average transmission fidelity $F_{n,d}$ given by Eq~(\ref{eq:sign:fidelity}) for undetected individual counterfeiting of $n$-qubit long messages extended by $d$ decoy qubits. The fidelity is averaged over all possible cases of successful counterfeiting of extended message which includes attacks on  $1$ to $n+d$ qubits.}
\end{figure}

\begin{figure}
\includegraphics[width=8cm]{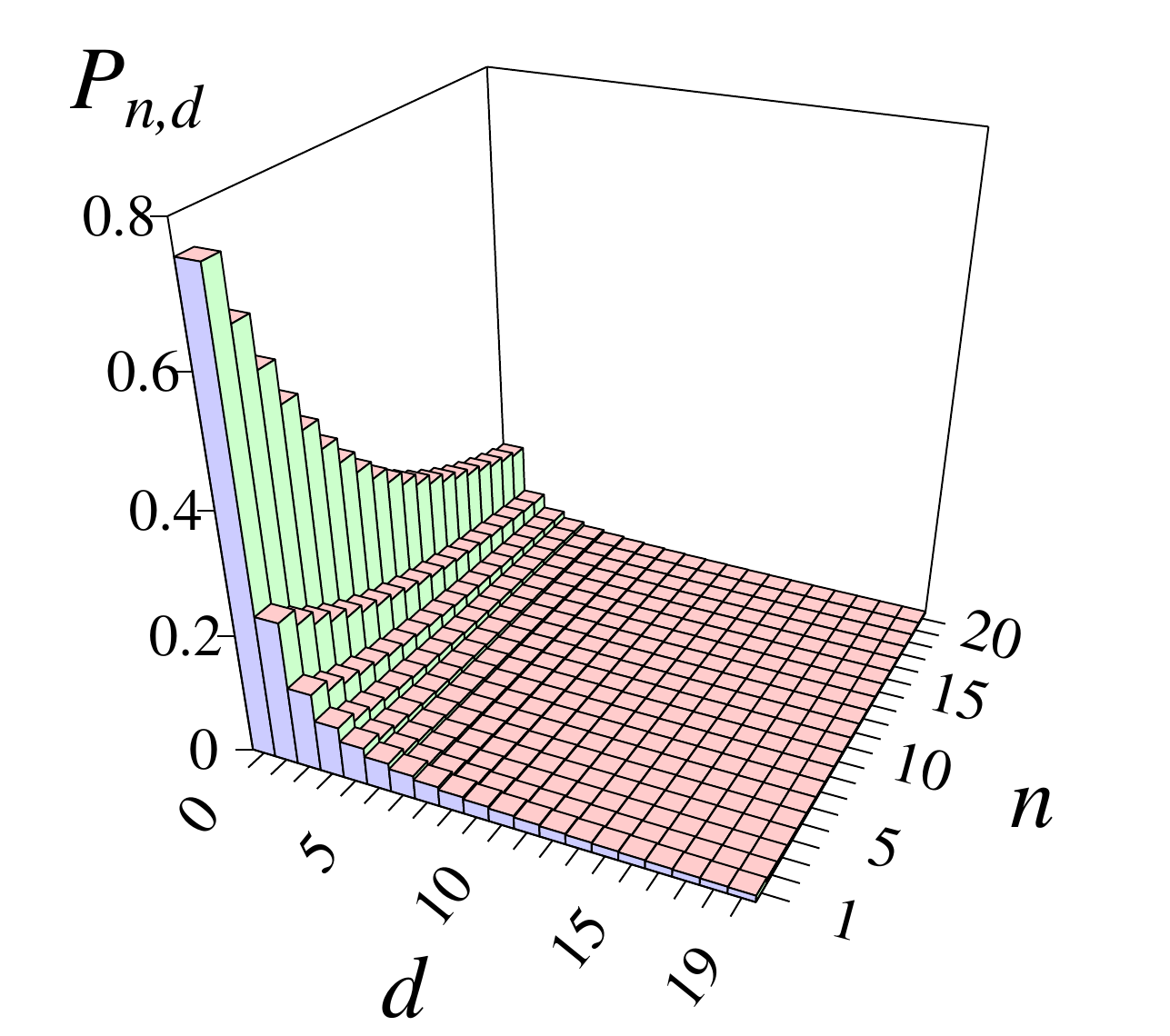}
\caption{\label{fig:Pnd} (color online) Same as in Fig.~\ref{fig:Fnd} but for the average counterfeiting probability $P_{n,d}$ given by Eq~(\ref{eq:PCa}).}
\end{figure}

It is also useful to investigate the average conditional fidelity of the transmitted chain of qubits, where  the condition is successful authentication.  Let us therefore consider an $n$-qubit long message with the addition of  $d$ decoy qubits, where  $k \leq n+d$ qubits are altered by Eve using the above described strategy. The unaltered qubits are transmitted with fidelity equal to $1$ while the counterfeited ones are transmitted with an average fidelity of $\frac{1}{2}$ (random states). As a result the fidelity of the transmission is given by 
\begin{equation}
F(k,m,n) = 1-\frac{k-m}{2n}
\end{equation}
and with probability $P_C(n,d,k,m)$ this transmission passes authenticity verification. Eve can attack an arbitrary number of qubits $k$ while successfully attaching $m$ decoy states. Thus, the scheme works on average with the mean fidelity
\begin{eqnarray}
\nonumber
 F_{n,d} &=& \frac{\sum\limits_{k=0}^{n+d} g(k)\sum\limits_{m=\mmin}^{\mmax}  P_C(n,k,m,d) F(k,m,n)}{\sum\limits_{k=0}^{n+d} g(k) \sum\limits_{m=\mmin}^{\mmax} P_C(n,d,k,m)} \\
&=& 1- \eta, \label{eq:sign:fidelity}
\end{eqnarray}
where $g(k)$ is a function describing the probability of Eve attacking $k$ qubits (for the sake of simplicity we assume that $g =1$, i.e., uniform distribution), $\mmin = \max[0,k-n]$, $\mmax = \min[d,k]$, and
\begin{eqnarray}
\label{eq:sign:eta}
\eta &=& \frac{\sum\limits_{k=0}^{n+d} \sum\limits_{m=\mmin}^{\mmax}  P_C(n,d,k,m) \frac{k-m}{2n}}{\sum\limits_{k=0}^{n+d} \sum\limits_{m=\mmin}^{\mmax} P_C(n,d,k,m)}.
\end{eqnarray}
Let us also define the average probability of counterfeiting as
\begin{equation}
\label{eq:PCa}
P_{n,d} = \frac{\sum\limits_{k=0}^{n+d} \sum\limits_{m=\mmin}^{\mmax} P_C(n,d,k,m)}{  \sum\limits_{k=0}^{n+d} \sum\limits_{m=\mmin}^{\mmax} }.
\end{equation}
An inspection of Eq.~(\ref{eq:sign:eta}) reveals that for the limit of long messages ($n\rightarrow\infty$), the value of $\eta$ becomes zero and the average fidelity reaches asymptotically the value of 1 (see Fig.~\ref{fig:Fnd}). Note, that by adding sufficient number of decoy states to the transmitted qubit stream one can asymptotically achieve  $\eta \to 0$. This is because, for a large number of decoy states $d\gg n$, it is more likely for decoy states to be attacked rather than the states forming the message [see Eq.~(\ref{eq:PC}) while ignoring the $3^k4^{-k}2^{-m}$ factor]. In other words, for large values of $d$ it is very likely that $k \approx m$. This means that the majority of errors will be found in the decoy states.  Secondly, for a large number of decoy states $d$, the probability of successful counterfeiting [see Eq.~(\ref{eq:PC}) while ignoring the probability of Eve attacking $m$ decoy states and $k-m$ message qubits] is significant only if $k\approx 1$. Thus, the mean over $k$ is negligible. The average probability of counterfeiting for a fixed value of $n$ drops rapidly with $d$ (see Fig.~\ref{fig:Pnd}).

So far we have worked under the assumption that Eve is unfamiliar with the content of the transmitted message. Let us now extend our analysis and assume that Eve managed to acquire complete information about the message and wishes to counterfeit it. Note that, since we are not dealing with cryptography but with authentication, the message itself is not a secret. Eve's strategy now consists of decoupling the message $|m\rangle$ from the key $|k\rangle$ qubits. At the time Eve still does not know the salt string, and so is unable to tell which qubit is the message and which is the key. She then picks one of the qubits at random and performs a projective measurement
\begin{equation}
\label{eq:sign:proj}
\Pi = |m\rangle\langle m| - |m^\perp\rangle\langle m^\perp|,
\end{equation}
where $|m^\perp\rangle$ is a qubit state orthogonal to $|m\rangle$. Whenever this projection measurement yields the value of $-1$, Eve discards the entire operation and the lost photons would appear to Bob as mere technological losses. There is a $\frac{1}{2}$ probability that Eve correctly picked the message qubit and subjected it to the projection measurement. In this case the measurement outcome is always $+1$ and Eve uses the remaining key qubit to authenticate a forged message qubit. When this happens, Bob does not register any errors ($\epsilon = 0$). On the other hand, there is a $\frac{1}{2}$ probability that Eve incorrectly picked the key qubit and subjected it to the projection measurement. Since there is no correlation between the key and message qubits, the measurement outcome would yield $+1$ or $-1$ with balanced probabilities. Considering that outcomes $-1$ are discarded, the overall probability of Eve sending an incorrect key is reduced by two. Assuming that, similarly to the previous analysis, Bob detects the incorrect key in half of the cases, the error rate introduced by Eve's wrong choice is $\epsilon = \frac{1}{4}$. In total, the average error rate (considering both good and bad guesses) introduced by Eve using this strategy is $\epsilon = \frac{1}{8}$ and  the probability of successfully counterfeiting $n$-qubit long message without being detected reads
\begin{equation}
P_C = \left(1-\epsilon\right)^n = \left(\frac{7}{8}\right)^n.
\end{equation}
Again, Alice can insert decoy qubits at random to arbitrarily decrease Eve's chances of undetected counterfeiting to 
\begin{equation}
P_C = \left(\frac{7}{8}\right)^n \left(\frac{1}{2}\right)^d,
\end{equation}
at the expense of lowering the transmission rate. 
The average fidelity in this scenario can be similarly analyzed. The result  would not differ quantitatively  from the ones in the preceding paragraph. Also note that Eve can adopt a strategy that is more complex than simple projection (\ref{eq:sign:proj}), but the analysis of this possibility is beyond the scope of this paper. However, for any individual attack strategy all the transmitted qubits can be attacked successfully  with probability $P_C$ proportional to $p_{\mathrm{decoy}}^d \approx 0$ for $d \gg 1$, where $p_{\mathrm{decoy}}<1$ is the probability of keeping a decoy state unaltered. This means that the malicious behavior is possible only if Eve knows the location of all the decoy states.

\subsection*{Adding noise to the message}
The security of this protocol depends on the symmetry between key and message qubits. Eve’s inability to decide which qubits belong to the key and which are the message makes it impossible for her to affect one without affecting the other. The message is not secret. However, if Eve tries to read it she will introduce errors in the key as well. A similar situation occurs when Eve decides to undertake a hostile action to discredit Alice by adding noise to her message without actually having any information about its content. In this scenario, Eve malevolently changes the message by applying a random transformation  unknown to Alice or Bob while the message itself is not known to Eve. Here we demonstrate that such action can be detected by Bob if he analyzes the key. Each of the qubits is assumed to be attacked independently. Even though we restrict our analysis to individual attacks, it is likely that the results hold even for collective attacks, since the message qubits are not correlated. Eve does not know if the message is represented by a spatial degree of freedom or the polarization state of a photon. In order to alter the message only, Eve has to guess the way in which its qubits are encoded. If she guesses correctly she will not be detected when changing the attacked qubit. She can choose to attack the spatial degree of freedom by tampering with one rail only, or she can attack the polarization degree of freedom by applying the same optical transformation in both rails. Both these strategies are equivalent. This is because Eve cannot always alter a message qubit $|m\rangle$ without being detected. If she attempts to change the qubit then, with probability  $\tfrac{1}{2}$, she will alter a key qubit $|k\rangle$ instead. Any single-qubit operation performed by Eve can be generated by Pauli's matrices $\sigma_n$ for $n=x,\,y,\,z$. Thus, if $|\langle \sigma_n \rangle_k | < 1$ for each $n$, an error can be detected at the last step of the protocol revealing Eve’s presence. The key qubit is given as 
\begin{equation}
|k\rangle = \tfrac{1}{\sqrt{2}}\left(|H\rangle \pm \mathrm{i}^r |V\rangle\right). 
\end{equation}
One can show that
\begin{equation}
|\langle \sigma_n \rangle_k |  = \left\lbrace
\begin{array}{cc}
\delta_{r,0}&\qquad\mbox{for}\qquad n = x\\
\delta_{r,1}&\qquad\mbox{for}\qquad n = y\\
0&\qquad\mbox{for}\qquad n = z
\end{array}\right.,
\end{equation}
where $\delta_{i,j}$ is Kronecker's delta. Alice alters the value of $x$ at random. This means that if Eve decides to perform $\sigma_x$ or $\sigma_y$ operation, her action will be detected with probability $\tfrac{1}{4}$. However, in case of $\sigma_z$ she will introduce errors with probability $\tfrac{1}{2}$. Hence, an eavesdropper cannot manipulate an individual message qubit without introducing errors that can be detected at the last step of the protocol. Note that there is some asymmetry between the $xy$-plane and the $z$ direction. This asymmetry could be removed by using all six eigenstates of Pauli's matrices as possible key qubits. However, as we have shown above, this is not necessary.  

The above analysis demonstrates the rationale behind using two sets of eigenstates of either  $\sigma_x$ or $\sigma_y$ as potential key qubits. If we decided to use only one basis, e.g., $r=0$, those  operations that can be described as rotations around the $x$-axis would not be detected. This would leave Bob unaware of Eve tampering with the quantum message. Another possible way around it would be to force Alice to send only eigenstates of $\sigma_x$ operator as a message. However, this would make the message classical and the protocol would be secure only with classical information.

According to the analysis presented in this subsection, Eve causes
authentication errors if she tries to malevolently forge the transmitted message even without actually knowing it or replacing it by her own message. Therefore, 
average fidelity can be calculated on the same premises, and with the same results as in Eq.~(\ref{eq:sign:fidelity}).
\section{Router assisted Bell state discrimination and non-destructive manipulation}
\label{sec:bell}
Bell state discrimination is a procedure used to project a two-qubit state onto one of the four maximally entangled Bell states
\begin{subequations}
\begin{eqnarray}
\label{eq:bell:output}
|\Phi^\pm\rangle & \equiv & \tfrac{1}{\sqrt{2}}\left(|00\rangle \pm |11\rangle\right),\\
|\Psi^\pm\rangle & \equiv & \tfrac{1}{\sqrt{2}}\left(|01\rangle \pm |10\rangle\right).
\end{eqnarray}
\end{subequations}
This procedure is { crucial to several quantum information algorithms,} including quantum teleportation \cite{bib:bouwmeester:teleport}, entanglement swapping \cite{zukowski93swapping} or dense coding \cite{benett92dense}. On the platform of linear optics, this task {is} problematic because of {the} probabilistic nature of single photon behavior on beam splitters.{A balanced} beam splitter and polarization projection can be used to discriminate {between} polarization encoded singlet state $|\Psi^-\rangle$ and one of the triplet states $|\Psi^+\rangle$ \cite{weinfurter94bs,braunstein95bs}. It would however be impossible to distinguish the remaining two triplet states $|\Phi^+\rangle$ and $|\Phi^-\rangle$ even if photon number resolving detectors are used. In general, deterministic and complete Bell state discrimination is impossible with linear optics \cite{lutkenhaus99perfectbell}. Therefore, one has to resort to either unambiguous protocols {go outside the scope} of linear optics \cite{calsamiglia01optimalbell}.

A number of {approaches to} full Bell state discrimination  {have} been published. For instance{,} additional photon ancillae can be used to perform this task  within  linear optics \cite{dusek01ancillae}. Alternatively, one can {consider more degrees of freedom} than polarization {by using hyperentanglement} \cite{wei07hyper}. An experimental demonstration of {this} procedure using polarization and orbital angular momentum has been presented in 2007 \cite{barbieri07hyper}. {A linear-optical} controlled-phase gate can serve as an probabilistic disentangling gate that transforms Bell states into mutually orthogonal and therefore distinguishable separable two-photon states \cite{bonato10cnot}. {It is also possible to} implement full Bell state discriminator using experimentally challenging non-linear optical phenomena \cite{he07nonlin}. {Also,} several schemes for {the} discrimination of higher dimensional entangled states have  been implemented \cite{schmid08higher}.

\begin{figure}
\includegraphics[scale=1]{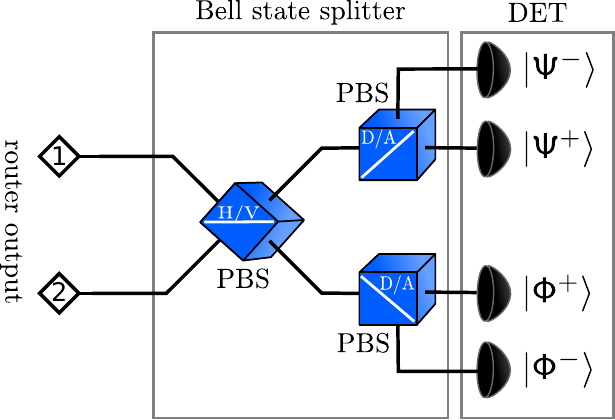}
\caption{\label{fig:bell:disc} (color online) Router assisted Bell state discriminator as described in the text{: PBS1, polarizing beam splitter transmitting horizontal and reflecting vertical polarization; PBS2 and PBS3,  beam splitters transmitting diagonal and reflecting anti-diagonal linear polarizations; DET, detection block with four detector each projecting onto one of the four Bell states.}}
\end{figure}
\begin{figure}
\includegraphics[scale=0.8]{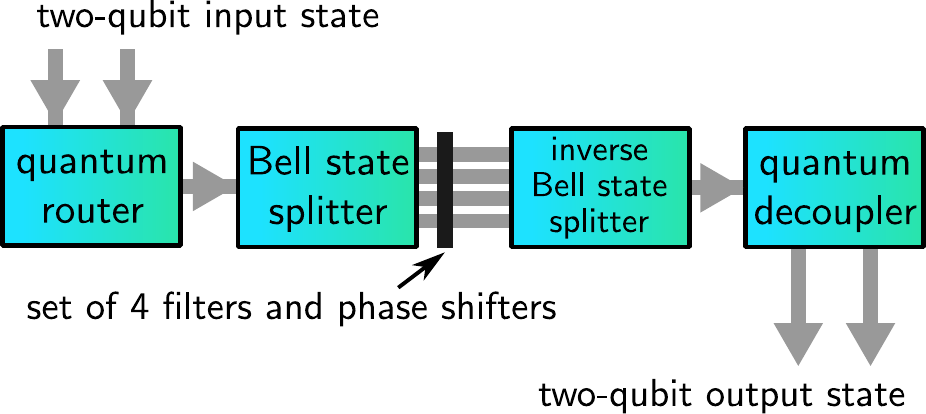}
\caption{\label{fig:bell:manip} (color online) Bell state manipulation using quantum router. The two-qubit state is processed by a quantum router. One of the qubits acts as signal,  the other as control. Subsequently, the output signal is split on Bell state splitter depicted in Fig. \ref{fig:bell:disc}. Set of four neutral density filters and phase shifters is used to {adjust} amplitude and phase of each individual Bell state separately. The signal is recombined on an inverse Bell state splitter  and finally two separate photons are recreated using the quantum decoupler described in Sec. \ref{fig:concept:decoupler}.}
\end{figure}
In this section we describe how a quantum router can assist in performing {a} complete Bell state discrimination. Furthermore, we extend our analysis to include complete non-destructive Bell state manipulation, allowing for coherent engineering of a two-qubit state directly in the Bell basis. In principle this task can not be achieved by any of the above mentioned techniques. So far {we have only considered  mutually separable signal and control states}. This assumption is valid if the router is a tool for steering the signal. {Nevertheless,} the router can deal with general signal and control input quantum state in the form of
\begin{equation}
\label{eq:bell:qubit}
|\psi_s\psi_c\rangle_{\mathrm{IN,C}} =c_{1} |00\rangle + c_{2} |01\rangle + c_{3} |10\rangle + c_{4} |11\rangle.
\end{equation}
Recalling the transformation implemented by the router (see Sec. \ref{ref:qr}), such input state gets transformed into
\begin{equation}
\label{eq:bell:qubitOUT}
|\psi_s\psi_c\rangle_{\mathrm{IN,C}}  \rightarrow c_{1} |0\rangle_1 + c_{2} |0\rangle_2 + c_{3} |1\rangle_1 + c_{4} |1\rangle_2,
\end{equation}
where lower indices denote the spatial mode of the output signal .

Let us now focus solely on the platform of polarization {qubits} and linear optics. The input state (\ref{eq:bell:qubit}) can be encoded using the horizontal and vertical polarization of individual photons
\begin{equation}
\label{eq:bell:photons}
|\psi_s\psi_c\rangle_{\mathrm{IN,C}}   = c_{1} |HH\rangle + c_{2} |HV\rangle + c_{3} |VH\rangle + c_{4} |VV\rangle.
\end{equation}
Alternatively, one can express the input state in terms of Bell states (Bell states basis)
\begin{equation}
\label{eq:bell:input}
|\psi_s\psi_c\rangle_{\mathrm{IN,C}}   = \alpha_1 |\Phi^-\rangle + \alpha_2 |\Phi^+\rangle + \alpha_3 |\Psi^-\rangle + \alpha_4 |\Psi^+\rangle.
\end{equation}
{It can be seen that the router processes} all Bell states individually, transcribing them into single-photon output state encoded in both polarization and spatial mode
\begin{subequations}
\begin{eqnarray}
\label{eq:bell:output}
|\Phi^\pm\rangle_{\mathrm{IN,C}} & \to & \tfrac{1}{\sqrt{2}}\left(|H\rangle_1 \pm |V\rangle_2\right),\\
|\Psi^\pm\rangle_{\mathrm{IN,C}} & \to & \tfrac{1}{\sqrt{2}}\left(|H\rangle_2 \pm |V\rangle_1\right),
\end{eqnarray}
\end{subequations}
where the lower indices denote spatial modes of the output . To distinguish between these four output states an additional setup (depicted in Fig.~\ref{fig:bell:disc}) is {added} to the output of the router. It consists of a Bell state splitter and {a} set of four single-photon detectors. In the first step, we separate the states {$|\Psi^\pm\rangle$} from $|\Phi^\pm\rangle$ by combining the two output modes on a polarizing beam splitter (transmitting horizontal and reflecting vertical polarization). {If the signal and control qubits of the router were in one of the $|\Psi^\pm\rangle$ ($|\Phi^\pm\rangle$) states, the photon from output of the router travels to the upper (lower) mode of the first PDBS  shown in Fig.~\ref{fig:bell:disc}. Hence, the combined transformation of the router followed by Bell splitting reads}
\begin{subequations}
\begin{eqnarray}
\label{eq:bell:outputPBS}
|\Phi^\pm\rangle_{\mathrm{IN,C}}   & \to & \tfrac{1}{\sqrt{2}}\left(|H\rangle \pm |V\rangle\right)_1 = |\pm\rangle_1,\\
|\Psi^\pm\rangle_{\mathrm{IN,C}}   & \to & \tfrac{1}{\sqrt{2}}\left(|H\rangle \pm |V\rangle\right)_2= |\pm\rangle_2.
\end{eqnarray}
\end{subequations}
In order to further distinguish between different Bell states we split each of the output modes in diagonal/anti-diagonal linear polarization basis. This can be performed by using a rotated polarizing beam splitter (transmitting diagonal and reflecting anti-diagonal linear polarization) or by implementing a Hadamard transform by a half-wave plate followed by a PBS splitting H and V polarization components. As depicted in Fig.~\ref{fig:bell:disc}, {full Bell state discrimination
is performed by detecting a photon by one of the four detectors which corresponds to a projection in Bell's basis.}

\subsection{Arbitrary Bell basis state manipulation}
In this subsection we explain how a quantum router can be used to perform arbitrary operations in Bell's basis. In contrast to Bell state discrimination, quantum state manipulation does not project the quantum state on one of the Bell states and, thus, maintains coherence. This task {cannot} be performed (at least not without restrictions) with just one controlled-phase gate. The controlled-phase gate transforms Bell states into separable mutually orthogonal two-photon states. These states remain correlated. It is, for instance, impossible to filter out just one of them from a two-photon state by applying local polarization-sensitive attenuation because a pair of states would be affected.

We have already described the {principles behind router-assisted} decomposition of a quantum state into Bell states. As illustrated in Fig.~\ref{fig:bell:manip}, the set of four detectors can be replaced by a set of four neutral density filters and phase shifters. {The} parameters of these filters and phase shifters can be set independently. {Hence, after applying inverse of the Bell splitter depicted in Fig. \ref{fig:bell:disc} followed by the quantum decoupler we can implement an arbitrary transformation}
\begin{subequations}
\begin{eqnarray}
|\Phi^-\rangle_{\mathrm{IN,C}}  & \to & \tau_1\mathrm{e}^{i\varphi_1}|\Phi^-\rangle_{\mathrm{S,A}} ,\\
|\Phi^+\rangle_{\mathrm{IN,C}}  & \to & \tau_2\mathrm{e}^{i\varphi_2}|\Phi^+\rangle_{\mathrm{S,A}} ,\\
|\Psi^-\rangle_{\mathrm{IN,C}}  & \to & \tau_3\mathrm{e}^{i\varphi_3}|\Psi^-\rangle_{\mathrm{S,A}} ,\\
|\Psi^+\rangle_{\mathrm{IN,C}}  & \to & \tau_4\mathrm{e}^{i\varphi_4}|\Psi^+\rangle_{\mathrm{S,A}} ,
\end{eqnarray}
\end{subequations}
where $\tau_m$ and $\varphi_m$ for $m=1,2,3,4$ are independent amplitude transmittances and phase shifts, respectively. Note that this procedure allows any two-qubit state in Bell basis to be directly engineered without disturbing its coherence. Using the above-described components one can implement an arbitrary positive valued measurement in the Bell's basis. This is especially interesting for investigating nonlinear properties of quantum states \cite{Bartkiewicz13correlations, Bartkiewicz13fid, Bartkiewicz13chsh}.    

\section{\label{sec:concl} Conclusions}
In this paper we have analyzed the properties of quantum routers and their corresponding inverse devices -- quantum decouplers. We have demonstrated how quantum routers can be used to implement prominent quantum information protocols on two examples: a quantum authentication procedure for the authentication of quantum messages, and the use of a quantum router for non-destructive complete Bell state manipulation. Both of these examples demonstrate that quantum routers should not be reduced to quantum analogues of classical routers, but rather considered to be important and versatile components of future quantum networks. The recent experimental demonstration of a linear-optical quantum router \cite{qfusion13} makes the theoretical concept of quantum routing experimentally feasible. This paper proposes viable options, which should be considered when dealing with practical quantum authentication or non-destructive Bell state manipulation.

\begin{acknowledgments}
K.~B. gratefully acknowledges the support by the Operational Program
Research and Development for Innovations -- European Regional
Development Fund (project CZ.1.05/2.1.00/03.0058 and the
Operational Program Education for Competitiveness - European
Social Fund (project CZ.1.07/2.3.00/20.0017 of the Ministry of
Education, Youth and Sports of the Czech Republic. 
K.~L. acknowledges the support by Czech Science 
Foundation (Grant no. 13-31000P). This work was
also supported by the Polish National
Science Centre under grant DEC-2011/03/B/ST2/01903.
\end{acknowledgments}


\begin{thebibliography}{99}

\bibitem{Nielsen_QCQI}  M.A. Nielsen and I.L. Chuang,   {\it Quantum Computation and Quantum Information},   Cambridge University Press, Cambridge, 2000.

\bibitem{Zeilinger_QIP}  D.~Bouwmeester, A.~Ekert, A.~Zeilinger:  {\it The Physics of Quantum Information},   Springer, Heidelberg, 2001.

\bibitem{bib:bruss:quantum_information} D.~Bruß and G.~Leuchs, {\em Lectures on Quantum Information}. \newblock Wiley-VCH, Berlin (2006).

\bibitem{gisin10ampl} N. Gisin, S. Pironio and N. Sangouard, \prl{\textbf{105}}, 070501 (2010).

\bibitem{Pitkanen11ampl} D. Pitkanen, X. Ma, R. Wickert, P. van~Loock, and N. L\"{u}tkenhaus, \pra{\textbf{84}}, 022325 (2011).

\bibitem{Curty11ampl} M. Curty and T. Moroder, \pra{\textbf{84}}, 010304(R) (2011).

\bibitem{micuda12ampl} M. Mičuda, I. Straka, M. Miková, M. Dušek, N. J. Cerf, J. Fiurášek, and M. Ježek,  Phys. Rev. Lett. {\bf 109}, 180503 (2012).

\bibitem{Kocsis13NPhys9} S. Kocsis, G.Y. Xiang, T.C. Ralph, and G.J. Pryde, Nat. Phys. {\bf 9}, 23--28 (2013).

\bibitem{bula13qnd} M.~Bula, K.~Bartkiewicz, A.~Černoch, K.~Lemr, Phys.~Rev.~A {\bf 87}, 033826 (2013).

\bibitem{scott13ampl} E.~Meyer-Scott, M.~Bula, K.~Bartkiewicz, A.~Černoch, J.~Soubusta, T.~Jennewein, K.~Lemr, Phys.~Rev.~A {\bf 88}, 012327 (2013).

\bibitem{Gisin02crypto} N. Gisin, G. Ribordy, W. Tittel, and H. Zbinden,  Rev. Mod. Phys. {\bf 74}, 145--195 (2002).

\bibitem{Bartkiewicz13}  K. Bartkiewicz, K. Lemr, A. \v{C}ernoch, J. Soubusta, and A. Miranowicz, \prl{\textbf{110}}, 173601 (2013). 
 
\bibitem{Ursin07QKD} R. Ursin \etal, {Nature Phys. {\bf 3}, 481 (2007)}.

\bibitem{Wang12QKD-260km} S. Wang \etal, Opt. Lett. {\bf 37} 1008 (2012).

\bibitem{Makarov10Hacking} L. Lydersen, C. Wiechers, C. Wittmann, D. Elser, J. Skaar, and V. Makarov, Nat. Photon. {\bf 4}, 686--689 (2010).

\bibitem{Lo12MDI-QKD} Hoi-Kwong Lo, Marcos Curty, and Bing Qi, \prl{\textbf{108}}, 130503 (2012).

\bibitem{shor95error} P.~W.~Shor, Phys.~Rev.~A {\bf 52}, R2493--R2496 (1995).

\bibitem{calderbank96error} A.~R.~Calderbank, P.~W.~Shor, Phys.~Rev.~A {\bf 54}, 1098--1105 (1996).

\bibitem{steane96error} A.~M.~Steane, Phys.~Rev.~Lett. {\bf 77}, 793--797 (1996).

\bibitem{gottesman98error} D.~Gottesman, Phys.~Rev.~A {\bf 57}, 127--137 (1998).

\bibitem{modlawska:non_maxi_entangl} J.~Modlawska and A.~Grudka,  Phys.~Rev.~Lett. {\bf 100}  (11),  110503  (2008).
  
\bibitem{paetznick13error} A.~Paetznick, B.~W.~Reichardt, Phys.~Rev.~Lett. {\bf 111}, 090505 (2013).

\bibitem{weinstein13error} Y.~S.~Weinstein, Phys.~Rev.~A {\bf 88}, 012325 (2013).

\bibitem{Hall11}   M.~A.~Hall, J.~B.~Altepeter, and P.~Kumar,  \prl~\textbf{106}, 053901 (2011).

\bibitem{lemr12router}  K. Lemr, A. Černoch,   Opt. Comm {\bf 300}, 282--285 (2013). 
  
\bibitem{lemr13router} K.~Lemr, K.~Bartkiewicz, A.~Černoch, J.~Soubusta, Phys.~Rev.~A {\bf 87}, 062333 (2013).
 
\bibitem{qfusion13} C.~Vitelli, N.~Spagnolo, L.~Aparo, F.~Sciarrino, E.~Santamato, L.~Marrucci, Nat. Photon. \textbf{7}, 521 (2013).

\bibitem{Zueco09}   D.~Zueco, F.~Galve, S.~Kohler, and P.~Hänggi,  \pra~\textbf{80}, 042303 (2009).

\bibitem{Aoki09}   T. Aoki \etal,  \prl~\textbf{102}, 083601 (2009).

\bibitem{Hoi11}   Io-Chun Hoi, C.M. Wilson, G. Johansson, T. Palomaki, B. Peropadre, and P. Delsing,  \prl~\textbf{107}, 073601 (2011).

\bibitem{liao10router} J.~-Q.~Liao, Z.~R.~Gong, L.~Zhou, Y.~-Xi.~Liu, C.~P.~Sun, F.~Nori, Phys.~Rev.~A {\bf 81}, 042304 (2010).

\bibitem{zhou13router} L.~Zhou, L.~-P.~Yang, Y.~Li, C.~P.~Sun, Phys.~Rev.~Lett. {\bf 111}, 103604 (2013).

\bibitem{lu14router} J.~Lu, L.~Zhou, L.-M.~Kuang, and F.~Nori, Phys.~Rev.~A {\bf 89}, 013805 (2014).

\bibitem{halenkova12detector} E.~Halenkov{\'a}, A.~\v{C}ernoch, K.~Lemr, J.~Soubusta, and S.~Drusov{\'a},   Appl. Opt. {\bf 51}  (4),  474--478  (2012).

\bibitem{kiesel05cphase} N.~Kiesel, C.~Schmid, U.~Weber, R.~Ursin, H.~Weinfurter, Phys.~Rev.~Lett. {\bf 95}, 210505 (2005).

\bibitem{Bartkiewicz13amp} K.~Bartkiewcz,  A. Černoch, K. Lemr, \pra{} \textbf{88}, 062304 (2013).

\bibitem{rivest90} R.~Rivest, {\em Cryptography} (Elsevier, 1990).

\bibitem{chuang2002patent} I.~Chuang, D.~Gottesman, patents US2002199108-A1, US7246240-B2 (2002).

\bibitem{chuang2001sign} I.~Chuang, D.~Gottesman, arXiv:quant-ph/0105032 (2001).

\bibitem{barnum02existence} H.~Barnum, C.~Crepeau, D.~Gottesman, A.~Smith, A.~Tapp, in {\em Proceedings of the 43th Annual IEEE Symposium on Foundations of Computer Science}, 449--458 (2002).

\bibitem{xiao06notrusted} W.~Xiao-jun, L.~Yun, arXiv:quant-ph/0509129 (2006).

\bibitem{wang06single}  J.~Wang, Q.~Zhang, C.~Tang, Optoel. Lett. {\bf 2}, 209--212 (2006).

\bibitem{sousa06contract} P.~B.~M.~de Sousa, R.~V.~Ramos, arXiv:quant-ph/0608232 (2006).

\bibitem{li13quantum} Q.~Li, W.~H.~Chan, C.~Wu, Z.~Wen, Intl. J. of Theor. Phys. {\bf 52}, 4335-4341 (2013).

\bibitem{lu05fourier} X.~Lu, D.~G.~Feng, in {\em Proceedings of the 7th International Conference on Advanced Communication Technology}, 514--517 (2005).

\bibitem{gao11arbitrated} F.~Gao, S.~Qin, F.~Guo, and Q.~Wen, Phys.~Rev.~A {\bf 84}, 022344 (2011).

\bibitem{zou10analysis} X.~Zou, D.~Qiu, Phys.~Rev.~A~{\bf 82}, 042325 (2010).

\bibitem{li09arbitrBell} Q.~Li, W.~H.~Chan, and D.-Y.~Long, Phys.~Rev.~A {\bf 79}, 054307 (2009).

\bibitem{zeng02arbitrated} G.~Zeng,C.~H.~Keitel, Phys.~Rev.~A {\bf 65}, 042312 (2002).

\bibitem{curty08optim} M.~Curty, N.~Lütkenhaus, Phys.~Rev.~A {\bf 77}, 046301 (2008).

\bibitem{zeng08arbutratedqs} G.~H.~Zeng, Phys.~Rev.~A {\bf 78}, 016301 (2008).

\bibitem{zou10qs} X.~F.~Zou, D.~W.~Qiu, Phys.~Rev.~A {\bf 82}, 042325 (2010).

\bibitem{curty02qubitauth} M.~Curty, D.~J.~Santos, and E.~Pérez, Phys.~Rev.~A~{\bf 66}, 022301 (2002).

\bibitem{dusek99indent} M.~Dusek, O.~Haderka, M.~Hendrych, and R.~Myska, Phys.~Rev.~A {\bf 60}, 149--156 (1999).

\bibitem{buhrman01finger} H.~Buhrman, R.~Cleve, J.~Watrous, and R.~de Wolf, Phys.~Rev.~Lett. {\bf 87}, 167902 (2001).

\bibitem{clarke12exper} P.~J.~Clarke, R.~J.~Collins, V.~Dunjko, E.~Andersson, J.~Jeffers, G.~S.~Buller, Nat. Communications {\bf 3}, 1174 (2012).

\bibitem{collins13exper} R.~J.~Collins, R.~J.~Donaldson, V.~Dunjko, P.~Wallden, P.~J.~Clarke, E.~Andersson, J.~Jeffers and G.~S.~Buller, arXiv:1311.5760 (2013).

\bibitem{assis12sign} F.~M.~Assis, A.~Stojanovic, P.~Mateus, Y.~Omar, Entropy {\bf 14}, 2531--2549 (2012).

\bibitem{bib:bouwmeester:teleport} D.~Bouwmeester, J.~Pan, K.~Mattle, M.~Eibl, H.~Weinfurter, and A.~Zeilinger, Nature (London) {\bf 390},  575  (1997).

\bibitem{zukowski93swapping} M.~Zukowski, A.~Zeilinger, M.~A.~Horne, A.~Ekert, Phys.~Rev.~Lett. {\bf 71}, 4287 (1993).

\bibitem{benett92dense} C.~H.~Bennett, S.~J.~Wiesner, Phys.~Rev.~Lett. {\bf 69}, 2881 (1992).

\bibitem{weinfurter94bs} H.~Weinfurter, Europhys.~Lett. {\bf 25}, 559 (1994).

\bibitem{braunstein95bs} S.~L.~Braunstein, A.~Mann, Phys.~Rev.~A {\bf 51}, R1727–R1730 (1995).

\bibitem{lutkenhaus99perfectbell} N.~Lütkenhaus, J.~Calsamiglia, K.-A.~Suominen, Phys.~Rev.~A {\bf 59}, 3295 (1999).

\bibitem{calsamiglia01optimalbell} J.~Calsamiglia, N.~Lütkenhaus, Appl.~Phys.~B {\bf 72}, 67--71 (2001).

\bibitem{dusek01ancillae} M.~Dušek, Opt. Comm. {\bf 199}, 161--166, (2001).

\bibitem{wei07hyper} T-C~Wei, J.~T.~Barreiro, P.~G.~Kwiat, Phys.~Rev.~A {\bf 75}, 060305(R) (2007).

\bibitem{barbieri07hyper} M.~Barbieri, G.~Vallone, P.~Mataloni, F.~De~Martini, Phys.~Rev.~A {\bf 75}, 042317 (2007).

\bibitem{bonato10cnot} C.~Bonato1, F.~Haupt, S.~S.~R.~Oemrawsingh, J.~Gudat, D.~Ding, M.~P.~van~Exter, D.~Bouwmeester, Phys.~Rev.~Lett. {\bf 104}, 160503 (2010).

\bibitem{he07nonlin} B.~He, J.~A.~Bergou, Y.~Ren, Phys. Rev. A {\bf 76}, 032301 (2007).

\bibitem{schmid08higher} C.~Schmid, N.~Kiesel, W.~Laskowski, W.~Wieczorek, M.~Żukowski, H.~Weinfurter, Phys.~Rev.~Lett. {\bf 100}, 200407 (2008).

\bibitem{Bartkiewicz13correlations} K.~Bartkiewicz, K.~Lemr, A.~Černoch,  J.~Soubusta, \pra{} \textbf{87}, 062102 (2013).

\bibitem{Bartkiewicz13fid} K.~Bartkiewicz, K.~Lemr, A.~Miranowicz, \pra{} \textbf{88}, 052104 (2013).

\bibitem{Bartkiewicz13chsh} K.~Bartkiewicz, B.~Horst, K.~Lemr, A.~Miranowicz, \pra{} \textbf{88}, 052105 (2013).

\end{thebibliography}
\end{document}